\documentclass[twocolumn,showpacs,prb,amsmath,amssymb]{revtex4}
\usepackage{graphicx}
\usepackage{subfigure}
\usepackage{bm}
\usepackage{color}
\usepackage{soul}

\newcommand{\beq}   {\begin{equation}}
\newcommand{\eeq}   {\end{equation}}
\newcommand{\ba}   {\begin{eqnarray}}
\newcommand{\ea}   {\end{eqnarray}}

\DeclareMathOperator{\Max}{Max}

\begin{document}

\title{Spin versus charge noise from Kondo traps}
\author{Luis G.\ G.\ V.\ Dias da Silva}
\affiliation{Instituto de F\'{\i}sica, Universidade de S\~{a}o Paulo,
C.P.\ 66318, 05315--970 S\~{a}o Paulo, SP, Brazil}
\author{Rog\'{e}rio de Sousa}
\affiliation{Department of Physics and Astronomy, University of Victoria, Victoria, British Columbia V8W 2Y2, Canada}
\date{\today}
\begin{abstract}
  Magnetic and charge noise have common microscopic origin in solid
  state devices, as described by a universal electron trap model. In
  spite of this common origin, magnetic (spin) and charge noise
  spectral densities display remarkably different behaviours when
  many-particle correlations are taken into account, leading to the
  emergence of the Kondo effect. We derive exact frequency sum rules for trap noise, and perform numerical renormalization-group calculations to show that while spin noise is a universal function of the Kondo temperature, charge noise remains well described by single-particle theory even when the trap is deep in the Kondo regime. We obtain simple analytical  expressions for charge and spin noise that accounts for Kondo screening in all frequency and temperature regimes, enabling the study of the impact of disorder and the emergence of magnetic $1/f$ noise from Kondo traps.
We conclude that the difference between charge and spin noise survives even in the presence of disorder, showing that noise can be
more manageable in devices that are sensitive to magnetic (rather than charge) fluctuations and that the signature of the Kondo effect can be observed in spin noise spectroscopy experiments.
\end{abstract}


\maketitle

\section{Introduction}
\label{sec:Intro}

The tunneling of conduction electrons into local charge traps is a prevalent
phenomena in solid state physics. Traps can be realized by artificial
structures such as quantum dots,\cite{latta11} or by natural ``unwanted''
defects such as dangling bonds \cite{desousa07} and bound states in metal/oxide interfaces. \cite{choi09} It has long been recognized that trap fluctuation causes charge noise in electronic devices, with the signature of individual traps being observed with a Lorentzian $1/f^2$ noise spectral density in small structures \cite{ralls84, desousa05} and an ensemble of them causing $1/f$
noise in large structures. \cite{weissman88} Here we address the fundamental question of how the electron \textit{spin} alters trap noise.

One of the greatest developments of interacting electron physics was the discovery that a local trap interacting with a Fermi sea gives rise to the Kondo effect, the formation of a many-body singlet with conduction electron spins screening out the local trap spin. \cite{wilsonRMP} The signatures of the Kondo effect in transport phenomena are well studied, but key issues related to dynamics have only been addressed recently with the emergence of modern Numerical Renormalization Group (NRG) algorithms. \cite{Bulla:395:2008}  
It is particularly interesting to find out whether trap noise will impact devices that are sensitive to \emph{magnetic fluctuations as opposed to charge}, e.g. spin-based or spintronic devices, \cite{zutic04, diao11} in the same way that
it affects conventional charge-based devices. Recent measurements of intrinsic magnetic flux noise in superconducting quantum interference devices do indeed confirm that trap spin fluctuation is the dominant source of noise.\cite{faoro08,sendelbach08,lanting14} Moreover, novel developments in spin noise spectroscopy \cite{crooker04} open several possibilities for the
detection of correlated spin fluctuations in quantum dot systems.

Given these interesting prospects, the question that we address here is the qualitative difference between pure charge/spin noise of a ``Kondo trap'' interacting with a Fermi sea, which we define as a local charge trap in the Kondo regime. 

The interplay of Kondo physics and noise has been mostly explored in the context of transport through quantum-dot  systems, with the Kondo trap right inside the transport path. In this case trap charge and spin fluctuation are intertwined in a non-trivial way. Calculations of the shot noise and current noise in different set-ups such as single \cite{Meir:Phys.Rev.Lett.:88:116802:2002,Moca:Phys.Rev.B:83:201303:2011,mueller10,Mueller:Phys.Rev.B:245115:2013,Moca:Phys.Rev.B:89:155138:2014}
and double quantum dots\cite{Lopez:Phys.Rev.B:69:235305:2004,Kubo:Phys.Rev.B:83:115310:2011,Breyel:Phys.Rev.B:84:155305:2011}
  in the Kondo regime have been reported.
Much less studied is the role of the Kondo state in \textit{spin} noise. The case of spin-current noise was considered in Refs. \onlinecite{Moca:Phys.Rev.B:81:241305:2010} and \onlinecite{Moca:Phys.Rev.B:84:235441:2011}, and qualitative differences between spin-current and charge-current noise were found to exist. 

In this article, we show that focusing on \emph{pure spin/charge trap noise} (i.e., finite frequency trap occupation noise) 
allows for a different perspective on the problem of Kondo trap dynamics: it enables a clear separation between the contributions of single-particle excitations and the many-particle processes 
connected with the formation of the Kondo singlet state. Moreover, considering pure spin (charge) trap noise is important for describing transport  experiments with traps outside the transport channel. In this case, trap  fluctuations produce bias magnetic (electric) noise that in turn may dominate the spin-current (charge-current) noise.

Our article is organized as follows. In Section~\ref{sec:Model} we outline our model for pure spin/charge  trap noise, and establish its connection to the usual spin/charge  susceptibilities. We demonstrate six exact results: four sum rules and two Shiba relations. In Section~\ref{sec:HF} we describe our Hartree-Fock (HF) or mean-field approximation, that mainly accounts  for single-particle processes. In Section~\ref{sec:NRG} we present  our non-perturbative NRG calculations, which account for   single-particle and many-particle processes on the same  footing. The NRG results show that finite-frequency spin/charge   noise have quite distinct behaviors and are dominated by completely   different processes. In Section~\ref{sec:analyticspin} we use NRG  and the sum rules to obtain an analytic approximation to spin noise   in the Kondo regime, and in Section~\ref{sec:disorder} we use this   analytic approximation to study the interplay between disorder and   Kondo correlations in an ensemble of Kondo traps. We show that, in   the presence of disorder, the spin noise displays   a temperature-dependent $1/f$ noise that is qualitatively distinct  from the temperature-independent charge $1/f$ noise. Finally, Section~\ref{sec:Conclusion} presents our concluding remarks, with a discussion of the impact of our results in the effort to detect Kondo correlations in spin noise spectroscopy experiments, and our prediction of qualitatively different $1/f$ noise impacting spin-based and charge-based devices.

\section{Charge trap model and exact sum rules}
\label{sec:Model}

Our starting point is the Anderson model \cite{anderson61} for a
trapping-center interacting with a Fermi sea, 
\begin{equation}
H \!=\!{\cal H}_{\rm band}+{\cal H}_{\rm hyb}+{\cal H}_{\rm trap},
\label{h}
\end{equation}
with 
\begin{subequations}
\begin{eqnarray}
{\cal H}_{\rm band}  &=&\sum_{k,\sigma}\epsilon_{k\sigma}n_{k\sigma},\\
{\cal H}_{\rm hyb} &=& \sum_{k,\sigma} V_{dk}\left(c^{\dag}_{k\sigma}d_{\sigma}+d^{\dag}_{\sigma}c_{k\sigma}\right),\\
{\cal H}_{\rm trap} &=&\epsilon_d\left(n_{\uparrow}+n_{\downarrow}\right) +U n_{\uparrow}n_{\downarrow}.
\end{eqnarray}
\end{subequations}
In the above, $c^{\dag}_{k\sigma}$ ($c_{k\sigma}$) is a creation
(destruction) operator for a conduction electron with wavevector $k$
and spin $\sigma=\uparrow,\downarrow$,
$n_{k\sigma}=c^{\dag}_{k\sigma}c_{k\sigma}$ counts the number of band
electrons in state $k,\sigma$ with energy
$\epsilon_{k\sigma}$. Similarly, the operators $d^{\dag}_{\sigma}$ and
$d_{\sigma}$ create and destroy a trap electron with spin
$\sigma$, respectively, with $n_{\sigma}=d^{\dag}_{\sigma}d_{\sigma}$
being the number operator for electrons with spin $\sigma$ occupying
the trap state with energy $\epsilon_d$. Finally, $U$ is the
Coulomb repulsion energy for the trap, with $\epsilon_d+U$ the
energy required to add a second electron to an trap site that already contains one electron. 

Our goal is to calculate the trap \textit{spin} $S_s(\omega, T)$ and
\textit{charge} $S_c(\omega,T)$ noise spectral densities, defined by:
\begin{equation}
S_{i=s,c}(\omega,T) = \frac{1}{2\pi}\int_{-\infty}^{\infty} dt  \;\textrm{e}^{i\omega t}
\left\langle 
\delta{\cal \hat{O}}_i(t)\delta{\cal \hat{O}}_i(0)\right\rangle,
\label{Eq:sz_ndnoise}
\end{equation}
where
$\delta{\cal \hat{O}}_i(t)={\cal \hat{O}}_i(t)-\langle{\cal
  \hat{O}}_i\rangle$
with trap spin and charge operators given by
${\cal \hat{O}}_s=S_z=(n_{\uparrow}-n_{\downarrow})/2$ and
${\cal \hat{O}}_c=(n_{\uparrow}+n_{\downarrow})$, respectively, and
$\langle \cdot\rangle$ denoting the thermal equilibrium average.

We write an exact expression for the spin and
charge noise by performing a spectral decomposition of Eq.~(\ref{Eq:sz_ndnoise}) in the basis of energy eigenstates:
\begin{equation}
S_{i}(\omega)= \sum_{m,n} \frac{\textrm{e}^{-E_{m}/T}}{Z} \left|\left\langle n|{\cal \hat{O}}_i|m \right\rangle
\right|^{2} \delta(\omega - E_{nm}) - \langle {\cal \hat{O}}_i \rangle^2 \delta(\omega) \;,
\label{Eq:noise_Lehmann_delta}
\end{equation}
where $Z$ is the partition function,  $|m \rangle$ are (many-body) eigenstates of the Hamiltonian (\ref{h}) with energy $E_{m}$ ($E_{nm}\equiv E_{n}-E_{m}$) and $\langle n|{\cal \hat{O}}_i|m \rangle$ are the many-body matrix elements of the local operator ${\cal \hat{O}}_i$. For simplicity, we set $\hbar=k_B=1$. 
Note that Eq.~(\ref{Eq:noise_Lehmann_delta}) implies that $S_{i}(\omega,T)\geq 0$ and $S_{i}(-\omega,T)=\textrm{e}^{-\omega/T}S_{i}(\omega,T)$ as required by our assumption of thermal equilibrium.

The noise spectra is closely related to the dynamical susceptibility
associated with the operator ${\cal \hat{O}}_i$. We shall explore this connection in order to derive the exact frequency sum rules and Shiba relations\cite{shiba75} for
$S_i(\omega,T)$. 
These relationships will be used in Sec.~\ref{sec:analyticspin} to obtain analytical approximations for the noise spectra.

Assuming that an external field $F_i(t)$ couples to ${\cal \hat{O}}_i$ through
${\cal H}_{{\rm ext}}=-{\cal \hat{O}}_iF_i(t)$, 
the linear response of ${\cal \hat{O}}_i$ to $F_i$ will be
$\langle {\cal \hat{O}}_i(t)\rangle_{F\neq 0} - \langle {\cal
  \hat{O}}_i\rangle_{F=0}= 2\pi \int d\omega \textrm{e}^{-i\omega t}\chi_{i}(\omega,T)F_i(\omega)$,
where $\chi_{i}(\omega,T)$ is the dynamical susceptibility given by \cite{kubo91}
\begin{equation}
\chi_{i}(\omega,T)=\frac{i}{2\pi}\int_{0}^{\infty}dt\;\textrm{e}^{i\omega t}\left\langle \left[{\cal \hat{O}}_i(t),{\cal \hat{O}}_i(0)\right]\right\rangle.
\label{chio}
\end{equation}

Performing a spectral decomposition of Eq.~(\ref{chio}) and comparing to Eq.~(\ref{Eq:noise_Lehmann_delta}) leads to the following Lehmann representation:
\begin{equation}
\chi_{i}(\omega,T)=\frac{1}{2\pi}\int_{-\infty}^{\infty}\frac{d\omega'}{\omega-\omega'+i\eta}\left[S_{i}(-\omega',T)-S_{i}(\omega',T)\right],
\label{lehmannchi}
\end{equation}
with $\eta\rightarrow 0^+$. Separating the susceptibility into real and imaginary parts, $\chi_i=\chi_i' + i \chi_i''$, using $S_{i}(-\omega,T)=\textrm{e}^{-\omega/T}S_{i}(\omega,T)$,
and taking the imaginary part of Eq.~(\ref{lehmannchi}) leads to
\begin{equation}
\chi''_{i}(\omega,T)=\frac{1-\textrm{e}^{-\omega/T}}{2}S_{i}(\omega,T),
\label{fttheorem}
\end{equation}
which is known as the fluctuation-dissipation theorem. Moreover, 
taking the real part of Eq.~(\ref{lehmannchi}) yields
\begin{equation}
\chi'_i(\omega,T) = \frac{1}{2\pi}{\cal P} \int_{-\infty}^{\infty} \frac{d\omega'}{\omega'-\omega}\left(1-\textrm{e}^{-\omega'/T}\right)S_i(\omega',T),
\label{kramerskronig}
\end{equation}
which is the Kramers-Kronig causality relation. 

We now derive the frequency sum rules. 
The first one is obtained by direct integration of Eq.~(\ref{Eq:noise_Lehmann_delta}) over all frequencies:
\begin{equation}
\int_{-\infty}^{\infty}S_{i}(\omega,T) \; d \omega  =   \langle {\cal
\hat{O}}^2_i \rangle - \langle {\cal \hat{O}}_i \rangle^2.
\label{sumrule1}
\end{equation}

We call this the \emph{spin} or the \emph{charge} sum rule depending on whether $i=s$ or $i=c$. 
Another sum rule is obtained by setting $\omega=0$ in Eq.~(\ref{kramerskronig}), and 
noting that Eq.~(\ref{lehmannchi}) implies $\chi_i(\omega=0,T)=\chi_i'(\omega=0,T)$:
\begin{equation}
\int_{-\infty}^{\infty}\frac{1-\textrm{e}^{-\omega'/T}}{2\pi \omega'}S_i(\omega,T)d\omega' = \chi_{i}(\omega=0,T).
\label{sumrule2}
\end{equation}
Accordingly, we call this the spin or charge \emph{susceptibility} sum rule. Altogether Eqs.~(\ref{sumrule1}),~(\ref{sumrule2}) form a set of four exact sum rules that are valid at any temperature $T$. 

Finally, there are two additional exact relationships between noise and susceptibility, that apply only at $T=0$. These are the so called Shiba relations:\cite{shiba75} 
\begin{subequations}
\begin{eqnarray}
{\rm Lim}_{\omega\rightarrow 0^+} \frac{S_s(\omega,T=0)}{8\pi^2\omega} &=& \left[\chi_s(\omega=0,T=0)\right]^{2},\label{shibaspin}\\
{\rm Lim}_{\omega\rightarrow 0^+} \frac{S_c(\omega,T=0)}{2\pi^2\omega} &=& \left[\chi_c(\omega=0,T=0)\right]^{2}.\label{shibacharge}
\end{eqnarray}
\end{subequations}
They imply that $S_i(\omega,T)$ is Ohmic (linear in
$\omega$) at $T=0$, with a slope related to the static susceptibility $\chi_i(\omega\!=\!0,T\!=\!0)$.

\begin{figure}
\includegraphics[width=0.5\textwidth]{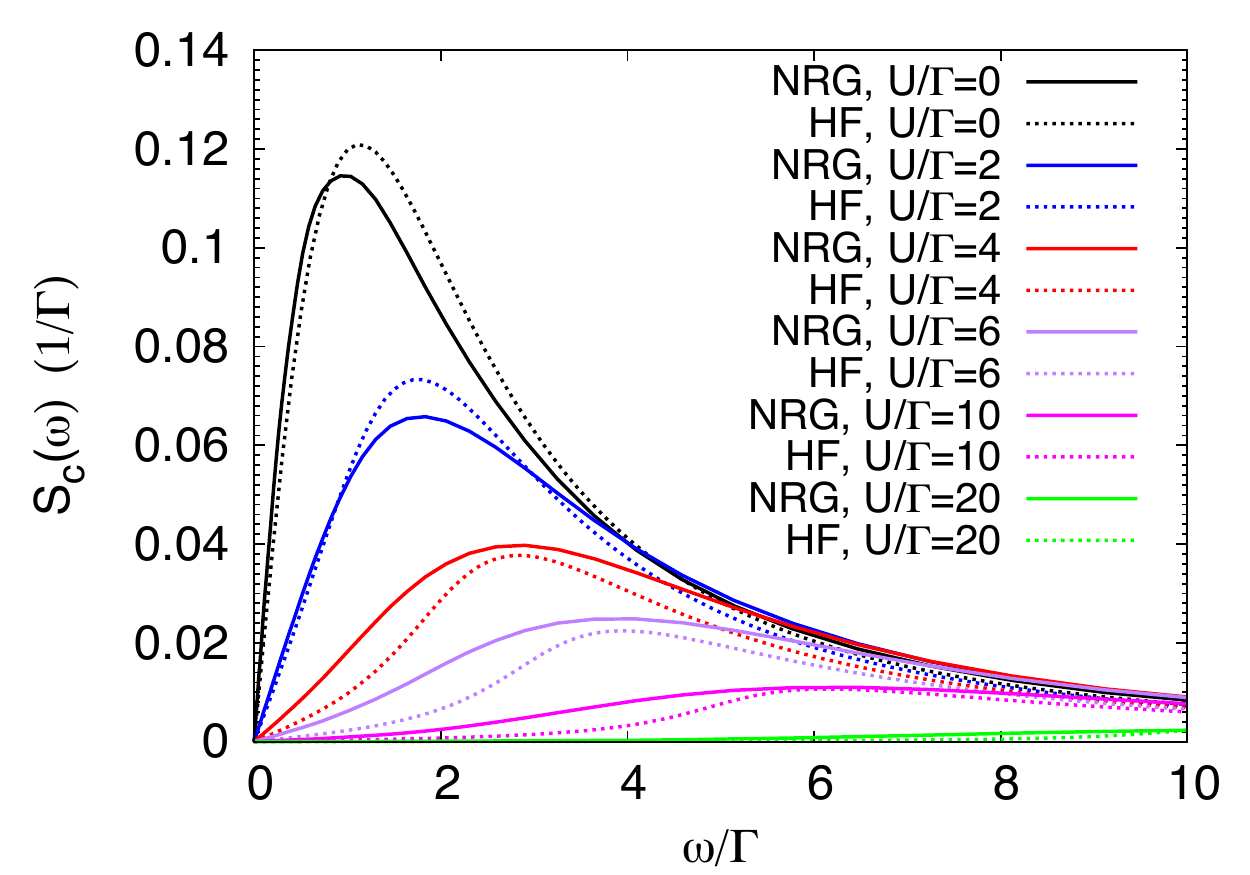}
\caption{(color online) Charge noise as a function of frequency for the trap in
the symmetric case with $\epsilon_d=-U/2$.
NRG calculations are shown to be  well approximated by
a mean-field Hartree-Fock
decomposition (HF) even when $U/\Gamma$ is large and the trap is deep in the
Kondo regime. This shows that charge noise is well described by single-particle
excitations.
}
\label{fig1}
\end{figure}

\section{Hartree-Fock approximation}
\label{sec:HF}

As a first approximation we calculate the noise spectral densities using
\emph{Hartree-Fock} (HF) decomposition based on writing expectation values into
products of spectral functions.\cite{anderson61} The advantage of HF is that
it becomes exact in the $U=0$ non-interacting limit.\cite{desousa05,desousa09} The
result for charge noise is
\begin{equation}
S_{c}^{{\rm HF}}(\omega,T)=\sum_{\sigma=\uparrow,\downarrow}\int d\epsilon A_{\sigma\sigma}(\epsilon)A_{\sigma\sigma}(\epsilon-\omega)[1-f(\epsilon)]f(\epsilon-\omega),
\label{mftc}
\end{equation}
and for the spin noise we get simply $S_{s}^{{\rm HF}}(\omega,T)=\frac{1}{4}S_{c}^{{\rm HF}}(\omega,T)$,
i.e., in the HF approximation magnetic noise is simply $\frac{1}{4}$ times the charge noise. In Eq.~(\ref{mftc}) $f(\epsilon)=1/[\exp{((\epsilon-\epsilon_F)/T)}+1]$ is the Fermi function, and
\begin{subequations}
\begin{eqnarray}
A_{\uparrow\uparrow}(\epsilon)&=&\frac{\Gamma/\pi}{(\epsilon-\epsilon_d)^{2}+\Gamma^{2}},\label{aup}\\
A_{\downarrow\downarrow}(\epsilon)&=&\frac{\Gamma/\pi}{(\epsilon-\epsilon_d-U)^{2}+\Gamma^{2}},\label{adown}
\end{eqnarray}
\end{subequations}
are HF local density of states for the trap with spin $\uparrow$ and
$\downarrow$, respectively.  The energy scale $\Gamma \equiv \pi \rho V^2_d$ models the rate for escape of a trap electron into the Fermi sea, with $\rho$ the energy density at the Fermi level, and $V_{dk} \equiv V_d$ a $k$-independent coupling between trap and Fermi sea. Note that Eqs.~(\ref{aup}) and ~(\ref{adown}) break the local spin symmetry by assuming the energy for the $\uparrow$ and $\downarrow$ trap states are $\epsilon_d$ and $\epsilon_d+U$,
respectively. This result is well known to be incorrect, in that it misses Kondo physics, i.e. the screening of trap spin by the electron gas spins.

\begin{figure}
\includegraphics[width=0.5\textwidth]{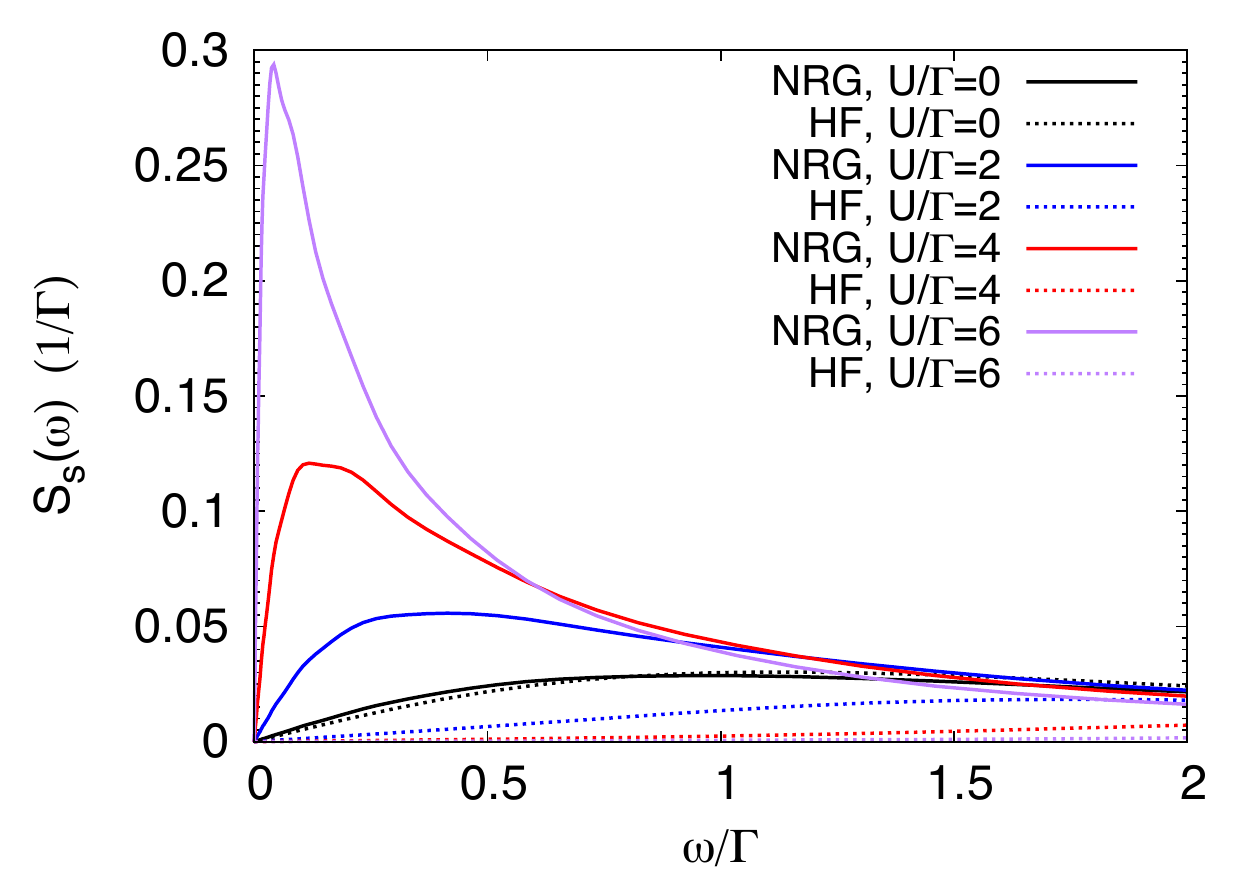}
\caption{(color online) Spin noise as a function of frequency for the trap in
the symmetric case with $\epsilon_d=-U/2$. The NRG results
agree with HF only
at $U=0$. As $U$ increases NRG shows that
the magnetic noise increases, developing
a peak at $\omega\approx T_K$. In contrast, the single particle contributions
described by HF
decrease dramatically as $U$ increases. This
shows that magnetic noise is dominated by many-body processes.
}
\label{fig2}
\end{figure}

\section{NRG calculations}
\label{sec:NRG}

We shall compare the Hartree-Fock approach to non-perturbative NRG calculations of the noise spectra, that take into account local spin symmetry and the formation of the Kondo singlet. The NRG algorithm calculates, within some well-controlled
approximations,\cite{Bulla:395:2008} the many-body spectrum for the Anderson model.\cite{Wilson1980,Bulla:395:2008} Conduction electrons are assumed to have a continuum spectrum, forming a metallic band with a half-bandwidth $D$.

At zero temperature, the first term in Eq.\
(\ref{Eq:noise_Lehmann_delta}) can be computed from the NRG spectral
data \cite{CostiHZ94,Bulla:045103:2001,Bulla:395:2008} down to
arbitrarily small non-zero frequencies $|\omega|>0$. The spectral
weight at $\omega\!=\!0$ and the fulfillment of the sum-rules can be
obtained by calculating the expectation values
$\langle {\cal \hat{O}}_i \rangle$ and
$\langle {\cal \hat{O}}^2_i \rangle$ with NRG. Since we will be
interested in the large frequency regime and our spectral functions
obey well-defined sum rules, we have chosen to use the ``Complete Fock Space" (CFS) approach \cite{Peters:Phys.Rev.B:245114:2006,
  Weichselbaum:Phys.Rev.Lett.:99:076402:2007} to calculate
$S_{i}(\omega>0)$ at zero temperature. As discussed in Appendix~\ref{sec:NRGdetails}, this choice has two important features: (i) the $T=0$ spectral functions are sum-rule-conserving by construction and (ii) broadening artifacts in the high frequency
regime, which can mask the correct power-law behavior, are
minimized. . 

\begin{figure}[t!]
\includegraphics[width=1.0\columnwidth]{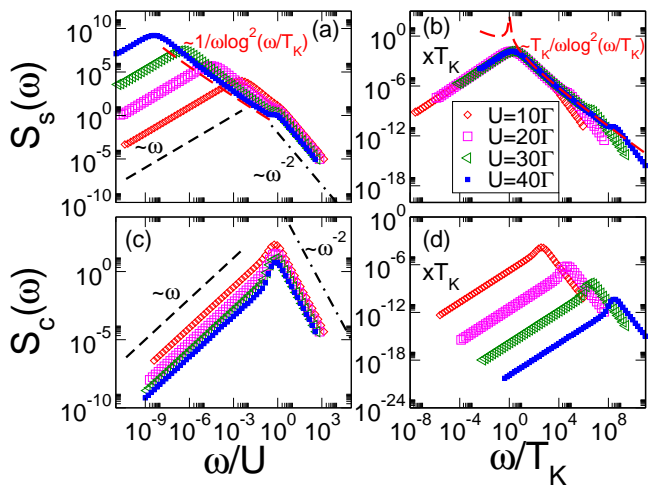}
\caption{(Color online) Universal scaling for spin noise in the Kondo regime
for $\epsilon_d=-U/2$. (a, b) NRG results for spin noise $S_s(\omega)$. Note
how all the curves collapse into a single scaling relation when the noise is
written as a function of $\omega/T_K$. For $\omega \lesssim  T_K$, the magnetic noise
scales linearly with  $\omega$ (Ohmic noise), and for $T_K \lesssim \omega < U$  it
decreases with an anomalous power of frequency $\propto
1/[\omega\log^{2}{(\omega/T_K)}]$. For $\omega > U$, spin noise  is cut-off
$\propto 1/\omega^{2}$. (c, d) NRG results for charge noise $S_c(\omega)$
\emph{do not show universal Kondo scaling}, and behaves just like the single
particle approximation (HF) with noise peaked at
$\omega\approx\Max{\{\Gamma,U\}}$
with smooth
cut-off $1/\omega^2$ at $\omega > U$.} \label{fig3}
\end{figure}

Figure~\ref{fig1} shows the calculated charge noise in the case
$\epsilon_d=-U/2$ for $\Gamma=10^{-4}D$ and
several different $U$. Remarkably, HF remains a good
approximation to charge noise even at large $U$. We interpret this result to be
evidence that charge noise is dominated by single particle processes \emph{even
when the trap is deep in the Kondo regime ($U\gg \Gamma$ for $T=0$)}.

The situation is drastically different for magnetic noise as shown in Fig.~\ref{fig2}. While NRG and HF agree with each other in the $U=0$ limit (when HF is exact), as soon as $U$ becomes non-zero the two methods show opposite results. As $U$ increases, the single particle noise (HF) decreases, while the many-body noise (NRG) increases. The low-frequency NRG results can be better visualized in Fig.~\ref{fig3}. We find (Fig.\ \ref{fig3}-a,c) that the
magnetic noise for a single trap is Ohmic at low frequencies, with a peak at
$\omega \approx T_K$ where $T_K$ is the Kondo temperature.
The magnetic noise spectral densities all collapse in the same universal curve
and scale as an anomalous power law $\propto
T_K/[\omega\log^{2}{(\omega/T_K)}]$ in the $T_K \!\ll\! \omega \!\ll\! U$
frequency range (Fig.~\ref{fig3}-b), consistent with previous results for the
 dynamical spin susceptibility 
\cite{KollerHM05,Garst:Phys.Rev.B:205125:2005,Glossop:Phys.Rev.B:104410:2007,fritsch10,Hoerig:Phys.Rev.B:165411:2014}
and the spin-current noise\cite{Moca:Phys.Rev.B:84:235441:2011}
in the Kondo regime.

\section{Analytical approximation for spin noise in the Kondo regime}
\label{sec:analyticspin}

While the HF approximation [Eq.~(\ref{mftc})] failed to describe spin
noise, it was shown to give a good description of charge noise at
$T=0$ [Fig.~\ref{fig1}]. In Appendix~\ref{sec:apphf} we show that \emph{HF
actually provides a good approximation for charge noise at
  $T\geq 0$}, in the sense that it approximately satisfies the sum rules and Shiba relations demonstrated in
Section~\ref{sec:Model}. The goal of the current section is to use our NRG calculations, 
sum rules and Shiba relations to obtain an analytical approximation
for spin noise at $T\geq 0$ in the Kondo regime.

It is well known\cite{Bulla:395:2008} that NRG has difficulty in calculating spectral features at frequencies $\omega < T$. Here, we propose an alternate approach to evaluate the spin noise for a broader $\omega/T$ range.

Motivated by the susceptibility sum rule Eq.~(\ref{sumrule2}) and the property  $S_{s}(-\omega,T)=\textrm{e}^{-\omega/T}S_{s}(\omega,T)$ we propose
the following fit function
\begin{equation}
S_{s}^{\rm Fit}(\omega,T) = \frac{2\omega\chi_{s}(\omega=0,T)}{1-\textrm{e}^{-\omega/T}}\frac{\Gamma_s}{\omega^{2}+\Gamma_s^{2}},
\label{ssfit}
\end{equation}
with the $\omega=0$ susceptibility given by a continuous fit to the NRG result\cite{hewsonbook}
\begin{widetext}
\begin{equation}
\chi_s(\omega=0,T)=\left\{
\begin{array}{c c}
\frac{{\cal W}}{8\pi T_K},& {\rm for}\;T\leq 0.23T_K,\\
\frac{0.68}{8\pi\left(T+\sqrt{2}T_K\right)}, & {\rm for}\; 0.23 T_K < T \leq 28.59 T_K,\\
\frac{1}{8\pi T}\left[1-\frac{1}{\log{\left(T/T_K\right)}}-\frac{\log{\left(\log\left(T/T_K\right)\right)}}{2\log^{2}{\left(T/T_K\right)}}\right], & {\rm for}\;T>28.59T_K,
\end{array}
\right.
\label{chi0T}
\end{equation}
\end{widetext}
where ${\cal W}=0.4128$ is the Wilson number.  

In Eq.~(\ref{ssfit}) $\Gamma_s\equiv\Gamma_s(\omega,T)$ is a fit function of frequency and temperature that will be determined by the exact sum rules and the Shiba relations. We recall that previous relaxational fits for $\Gamma_s$ assume no frequency dependence.\cite{Miranda:JournalofPhysics:CondensedMatter:9871:1996}
Here we allow $\Gamma_s(\omega,T)$ to vary on frequency so that the
logarithmic frequency decay discussed in Section~\ref{sec:NRG} is
properly accounted for.  

For $T\gg T_K$, the perturbative method of Suhl-Nagaoka\cite{gruner74,fritsch09} yields the high temperature
limit (the Korringa law):
\begin{equation}
\Gamma_s(\omega,T\gg T_K) \approx \frac{1}{4\pi}\frac{T}{1+\frac{4}{3\pi^{2}}\log^{2}{\left(\frac{T}{T_K}\right)}}.
\label{korringa}
\end{equation}

At $T=0$ the Shiba relation (\ref{shibaspin}) applied to Eq.~(\ref{ssfit})  implies\cite{Miranda:JournalofPhysics:CondensedMatter:9871:1996}
\begin{equation}
\Gamma_s(\omega=0,T=0)=\frac{1}{4\pi^2\chi_s(0,0)}=\frac{2T_K}{\pi {\cal W}},
\label{gammaszero}
\end{equation}
where we used the NRG result $\chi_s(0,0)={\cal W}/(8\pi T_K)$.

In order to interpolate between  Eqs.~(\ref{korringa})~and~(\ref{gammaszero}) we propose the following expression:
\begin{equation}
\Gamma_s(\omega,T)=\frac{1}{4\pi}
\frac{T+\frac{8}{{\cal W}}T_K}{1+\frac{1}{3\pi^2}
\log^{2}{\left[1+\left(\frac{T}{T_{K}}\right)^{2}+\left(\frac{\omega}{\alpha T_K}\right)^{2}\right]}},
\label{gammasfit}
\end{equation}
where $\alpha$ is a fit parameter to be determined by the spin sum rule [Eq.~(\ref{sumrule1})]:
\begin{equation}
{\rm Sum}_{s}(T)=4\int_{-\infty}^{\infty}d\omega S_{s}^{\rm Fit}(\omega,T).
\label{sumsT}
\end{equation}

This sum rule is most sensitive to $\alpha$ at $T=0$, and we find that the optimal fit value is quite close to $\alpha=3$, when ${\rm Sum}_s(T=0)=0.9994$. As an independent check, we evaluate the spin sum rule at $T>0$ and the susceptibility sum rule at $T\geq 0$:
\begin{equation}
{\rm Sum}_{\chi_s}(T)=\frac{1}{\chi_s(0,T)}\int_{-\infty}^{\infty}d\omega \frac{1-\textrm{e}^{-\omega/T}}{2\pi \omega}S_{s}^{\rm Fit}(\omega,T).
\label{sumchis}
\end{equation}

\begin{figure}[t]
\includegraphics[width=1.0\columnwidth]{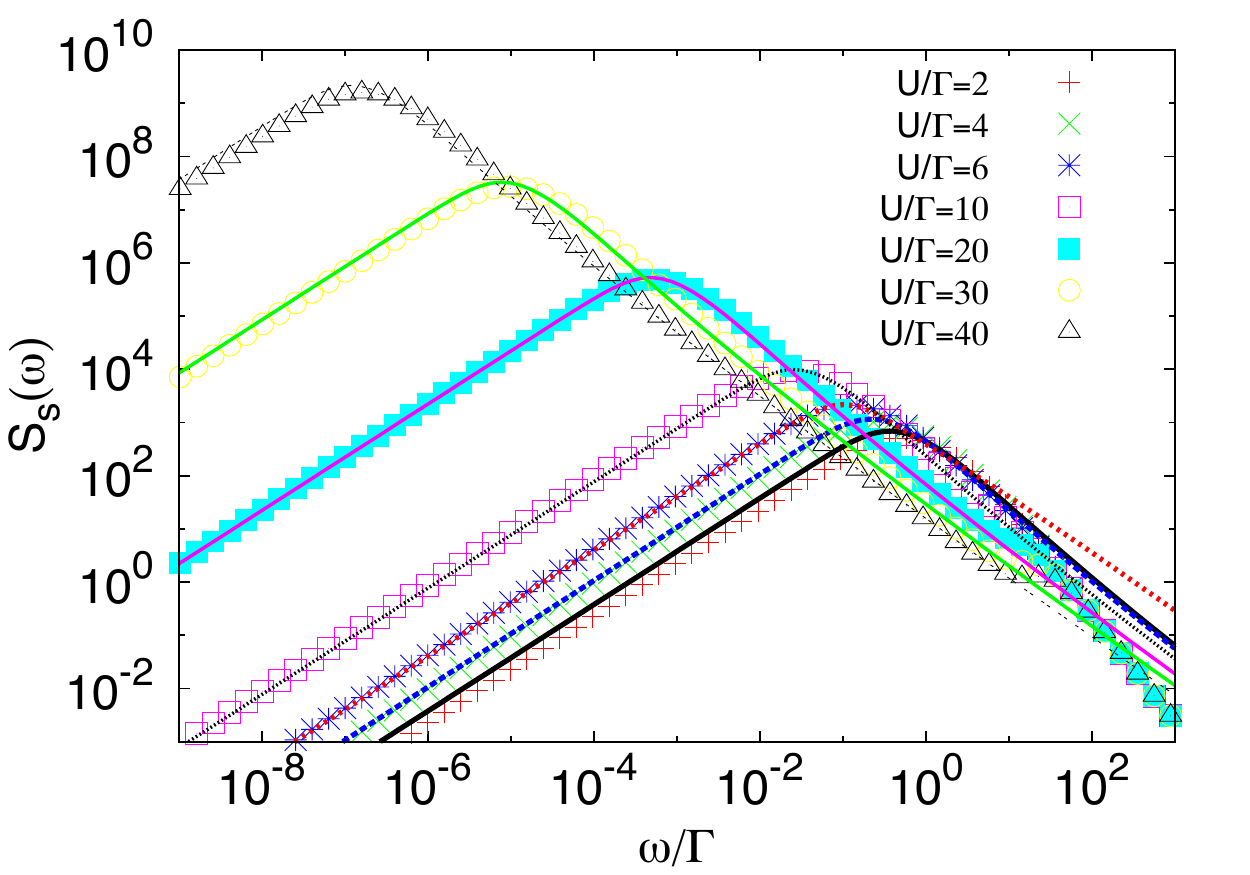}
\caption{ (Color online) Comparison of the spin noise fit $S^{\rm Fit}_s(\omega)$ [Eqs.~(\ref{ssfit})~and~(\ref{gammasfit}) with $\alpha=3$] (lines) with the NRG results (symbols) at $T\!=\!0$.
}
\label{fig:NewFitSpinNoise}
\end{figure}

In all cases, we obtain agreement within 36\%. A few examples are shown in Table~\ref{tablespincheck}. Moreover, we find that Eqs.~(\ref{ssfit})~and~(\ref{gammasfit}) with $\alpha=3$ provide an excellent fit of our NRG results at $T=0$, as shown in Fig.~\ref{fig:NewFitSpinNoise}.

Note that the choice of Eq.~(\ref{gammasfit}) implies  that $T_K S^{\rm Fit}_{s}(\omega,T)$ is a universal  function of $\omega/T_K$ and $T/T_K$, and that the presence of the temperature-dependent functions $\chi_s(0,T)$ and $\Gamma_s(\omega,T)$ suggest that spin noise has a much stronger temperature dependence than charge noise. 
In particular, Eq.~(\ref{ssfit}) fully accounts for the
Kondo screening for $T\!<\!T_K$ through $\chi_s(\omega\!=\!0,T)$.

\begin{figure*}
\centering
\subfigure[\;Low temperature behavior.\label{fig5a}]{\includegraphics[width=0.49\textwidth]{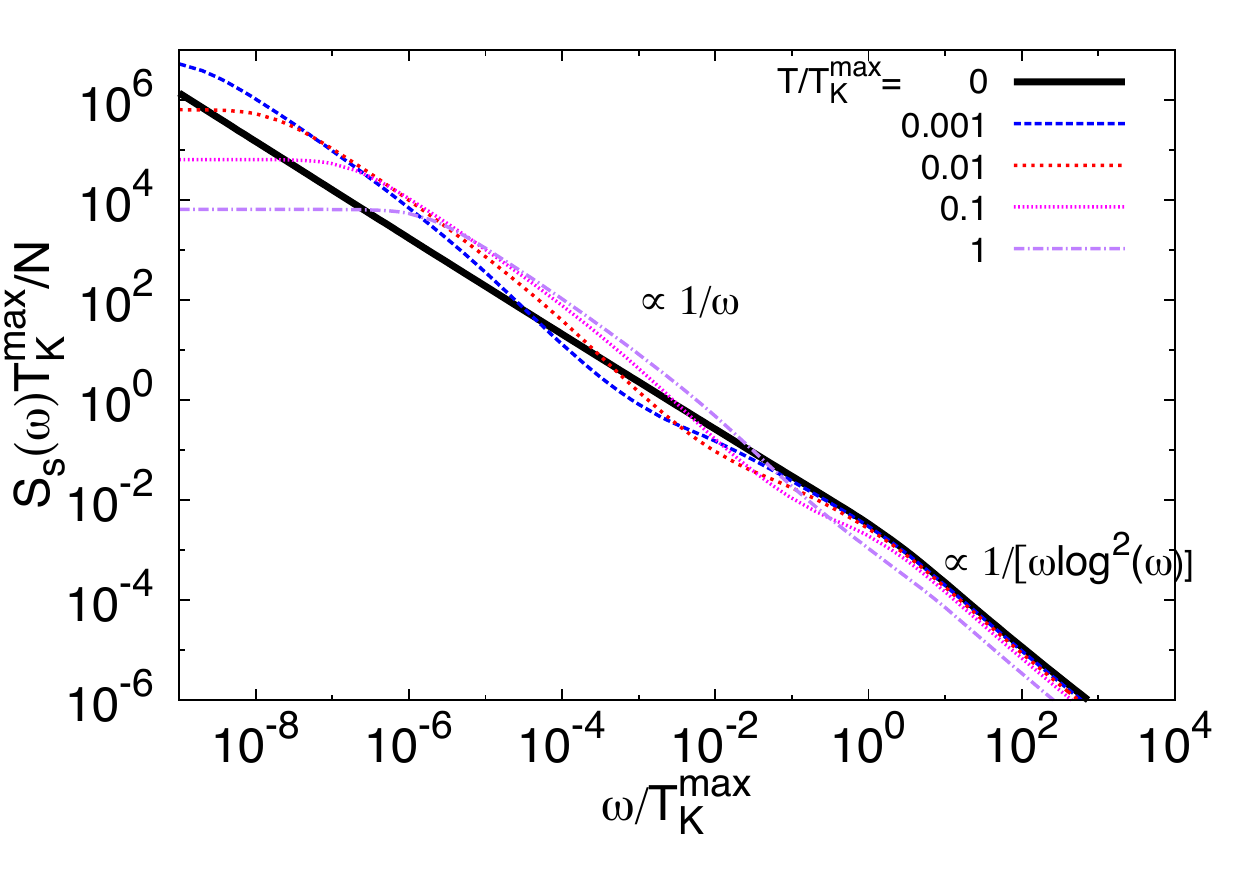}}
\subfigure[\;High temperature behavior.\label{fig5b}]{\includegraphics[width=0.49\textwidth]{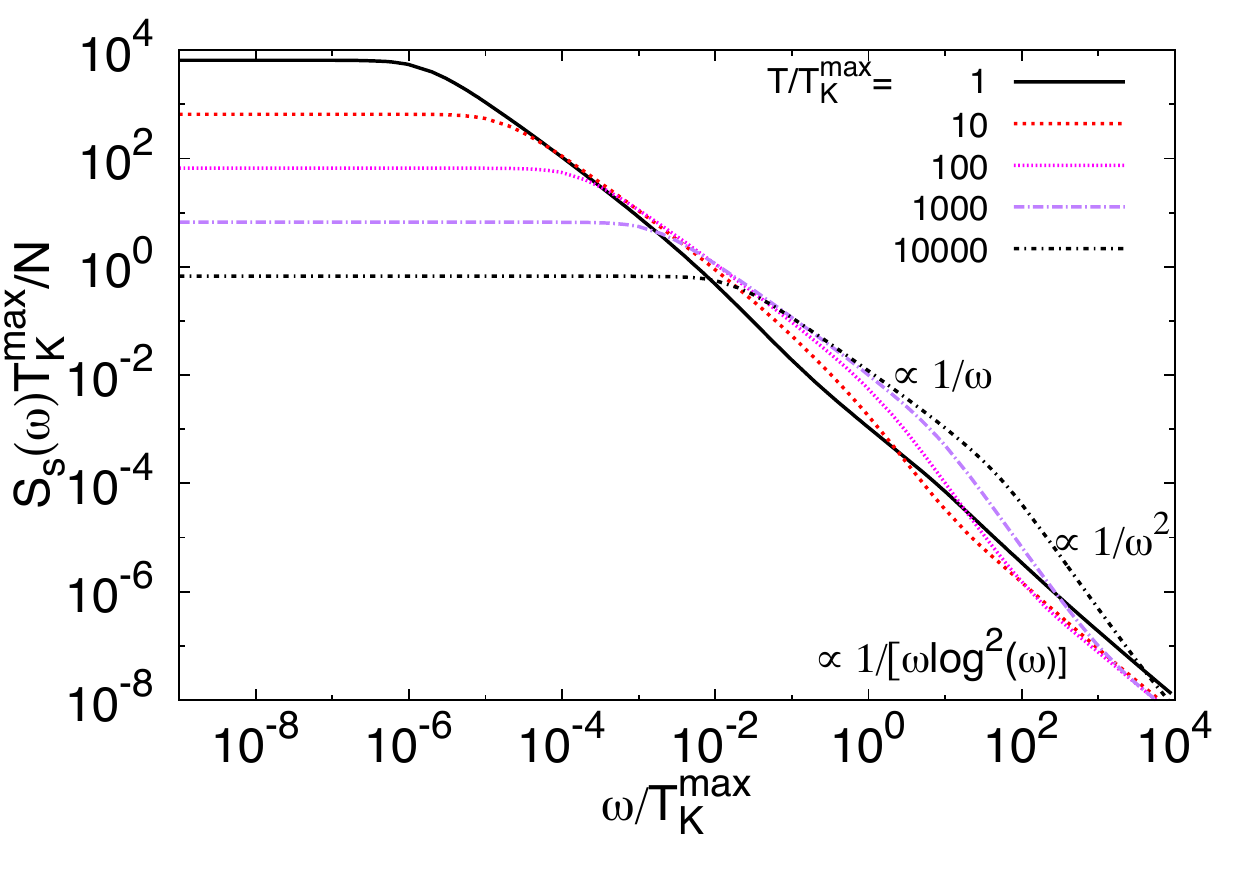}}
\caption[]{(Color online) Spin noise in the presence of trap disorder. The calculated noise for $N$ traps was averaged according to the prescription $\Gamma= \Gamma_0 \textrm{e}^{-\lambda}$, with $\lambda$ the tunneling distance between trap and the Fermi gas uniformly distributed in the interval
$[0,\lambda_{{\rm max}}]$. This gives rise to the broad distribution of Kondo temperatures shown in Eq.~(\ref{ptk}). The resulting noise, shown here for $\kappa=10$ and $\lambda_{{\rm max}}=5$ displays $1/f$ behavior over a frequency range that decreases as the temperature increases.  This is in contrast to the temperature independent
charge $1/f$ noise described in the literature.\cite{weissman88}}
\label{fig5}
\end{figure*}

\begin{table} 
\begin{center} 
\begin{tabular}{c |c |c } 
$T/T_K$ & ${\rm Sum}_s$ & ${\rm Sum}_{\chi_s}$\\
\hline 
$0$  &  $0.9994$ & $0.9518$\\ 
$0.5$  &  $0.9247$ & $0.9502$\\ 
$1$  &  $0.8245$ & $0.9503$\\ 
$10$  &  $0.6910$ & $0.9875$ \\ 
$100$  &  $0.7777$ & $0.9979$\\ 
\hline 
\end{tabular} 
\caption{Sum rules [Eqs.~(\ref{sumsT})~and~(\ref{sumchis})] applied to
  our analytical fit of spin noise, Eqs.~(\ref{ssfit}) and
  (\ref{gammasfit}) with $\alpha=3$. For the spin sum rules we used
  analytical approximations for $\chi_s(0,T)$ obtained by NRG
  [Eqs.~(4.53)~and~(4.60) in Ref.~\onlinecite{hewsonbook}]. In all
  cases we find that the sum rules are satisfied within 30\%.}
\label{tablespincheck} 
\end{center} 
\end{table} 

\section{Spin noise in the presence of disorder}
\label{sec:disorder}

In the case of an ensemble of $N$ Kondo traps, the noise will be
affected by disorder. The usual model for trap disorder (the one
that gives rise to ubiquitous charge $1/f$ noise)\cite{weissman88} is
to assume trap tunneling rate
$\Gamma = \Gamma_0 \textrm{e}^{-\lambda}$, where $\lambda$ models the
tunneling distance between trap and Fermi sea. The model assumes
$\lambda$ uniformly distributed with density
$P'(\lambda)=N/\lambda_{{\rm max}}$ for
$\lambda\in [0,\lambda_{{\rm max}}]$, and $P'(\lambda)=0$ for
$\lambda$ outside this interval, resulting in
$P(\Gamma)=(N/\lambda_{{\rm max}})/\Gamma$ and the corresponding
$1/f$ frequency dependence for trap charge noise. As we shall show, this same model
applied to Kondo traps gives rise to a much broader
distribution of Kondo temperatures that we denote $P(T_K)$.

For definiteness, we assume all Kondo traps have fixed $\epsilon_d$ and $U$, with the disorder solely affecting the parameter $\Gamma(\lambda)$. The dependence of the Kondo temperature
with $\lambda$ is given by\cite{haldane78}
\begin{eqnarray}
T_K(\lambda)&=&\sqrt{\frac{\Gamma(\lambda) U}{2\pi}}\textrm{e}^{\frac{\sqrt{3}\epsilon_d(\epsilon_d+U)}{U}\frac{1}{\Gamma(\lambda)}}\nonumber\\
&=&T_{K}^{{\rm max}}\textrm{e}^{-\left[\frac{\lambda}{2}+\kappa \left(\textrm{e}^{\lambda}-1\right)\right]}.
\end{eqnarray}
Here $\kappa=-\sqrt{3}\epsilon_d (\epsilon_d+U)/(U\Gamma_0)>0$ characterizes the type of trap. We shall assume $\kappa\gg (\lambda_{{\rm max}}+1)/2$, a limit that is typically satisfied by Kondo traps with $U\gg \Gamma$. The maximum and minimum Kondo temperatures of the distribution are given by
$T_{K}^{{\rm max}}=T_K(\lambda=0)$ and
$T_{K}^{{\rm min}}=T_K(\lambda=\lambda_{{\rm max}})$,
respectively; for $T_K\in [T_{K}^{{\rm min}},T_{K}^{{\rm max}}]$
the trap density becomes
\begin{equation}
P(T_K)=\frac{P'(\lambda)}{\left|\frac{dT_K}{d\lambda}\right|}
\approx \frac{\frac{N}{\lambda_{{\rm max}}} }{T_K\left[\kappa-\log{
\left(\frac{T_K}{T_{K}^{{\rm max}}}\right)}\right]},
\label{ptk}
\end{equation}
with $P(T_K)=0$ for
$T_K\not\in [T_{K}^{{\rm min}},T_{K}^{{\rm max}}]$. Note how $P(T_K)$
is exponentially broader than $P(\Gamma)$: we have
$T_{K}^{{\rm max}}/T_{K}^{{\rm min}}\approx
\exp{\left[\kappa\exp{\left(\lambda_{{\rm max}}\right)}\right]}$,
in contrast to
$\Gamma_{{\rm max}}/\Gamma_{{\rm min}}=\exp{\left(\lambda_{{\rm
        max}}\right)}$.
In spite of this difference, 
the normalization condition
$\int dT_K P(T_K)\approx N$ still holds since the logarithm in Eq.~(\ref{ptk}) makes $P(T_K)$ flatter than a $\sim 1/T_K$ distribution, thereby making the integral finite. We remark that our $P(T_K)$ is appropriate to describe highly disordered traps, such as traps randomly distributed at an insulator close to the metal/insulator interface. This situation is quite different from Kondo impurities in bulk alloys, whose $P(T_K)$ is considerably less
broad.\cite{Miranda:JournalofPhysics:CondensedMatter:9871:1996,bernal95}

Applying this averaging prescription to our spin noise Eq.~(\ref{ssfit}) yields
\begin{equation}
\left\langle S_{s}(\omega)\right\rangle = \int_{T_{K}^{{\rm min}}}^{T_{K}^{{\rm max}}} 
d T_K P(T_K) S_{s}^{\rm Fit}(\omega,T).
\end{equation}

The results are shown in Fig.~\ref{fig5}-a,b. 
At low temperatures ($T<T_{K}^{{\rm max}}$) the noise shows
$1/f$ behavior up to frequencies of the order of $T_{K}^{{\rm max}}$; at
larger frequencies, Kondo-enhanced exchange processes lead to a
$1/[f\log^{2}(f)]$ behavior.
For higher temperatures ($T>T_{K}^{{\rm max}}$) the noise saturates 
in the low frequency region, and the $1/[f\log^{2}(f)]$ behavior gets washed out of the high frequency region. 

Interestingly, the frequency range with $1/f$ behavior gets \textit{reduced} as the temperature increases. 
This shows that spin $1/f$ noise behavior is strongly temperature-dependent, in marked contrast to the usually temperature-independent charge $1/f$ noise.  The additional temperature dependence implies that temperature actually competes against disorder, converting the spin $1/f$ noise into a Lorentzian.

\section{Concluding Remarks}
\label{sec:Conclusion}

In conclusion, we presented a theory of charge and spin noise of a Kondo trap interacting with a Fermi sea. We showed that trap spin noise is qualitatively different from charge noise, in that the former occurs due to many-body scattering processes, while the latter is mainly dominated by single-particle
tunneling. This difference implies that spin noise has a stronger temperature
dependence than charge noise, and that it is controllable by tuning Kondo
temperature $T_K$ rather than trap tunneling rate $\Gamma$. 

Kondo trap dynamics displays two quite distinct behaviors depending on which property is probed. The experimental methods of charge
\cite{latta11} and spin \cite{crooker04} noise spectroscopy use
optical absorption to detect noise via the fluctuation-dissipation
theorem [optical absorption at frequency $\omega$ is
  directly proportional to $\chi_{i}''(\omega,T)$ and to noise as in
  Eq.~(\ref{fttheorem})]. Our results elucidate how Kondo
correlations can be observed with these methods. Pure charge
absorption does not enable the detection of the Kondo effect; in
Ref.~\onlinecite{latta11} the formation of the exciton state mixes
charge and spin fluctuation, and this feature was critical in enabling
their observation of the Kondo effect. For spin noise spectroscopy,
universal scaling with Ohmic behavior at
$T\!<\!\omega \!<\! T_K$ coupled with a
$1/[\omega\log^{2}{(\omega/T_K)}]$ tail for $T_K\!\ll\!\omega\!\ll\!U$
can be taken as the signature of the Kondo effect, allowing the
extension of this technique to probe Kondo correlations. However, in the presence of strong disorder over a range of Kondo temperatures $T_K\in [T_{K}^{{\rm min}},T_{K}^{{\rm max}}]$, we find that the Ohmic behavior is washed out, and the signature of Kondo correlations are visible only for $\omega > T_{K}^{{\rm max}}$ and
$T < 10 T_{K}^{{\rm max}}$ [See Fig.~\ref{fig5}b].

The qualitative difference between spin and charge noise survives even in the presence of disorder and high temperatures (namely $T_{K}\gg T_{K}^{{\rm max}}$).  As the temperature increases, the range of $1/f$ behavior for spin noise decreases, while the range of $1/f$ charge noise remains essentially unaltered. 
The additional temperature dependence for spin noise implies that temperature actually competes against disorder, converting the spin noise $1/f$ behavior into a Lorentzian-like dependence. Given that $1/f$ noise is notoriously difficult to control,\cite{paladino14} we reach the conclusion that ubiquitous trap noise can be more manageable in spin or flux-based devices that are sensitive to magnetic fluctuations rather than charge.

\begin{acknowledgements}
\emph{Acknowledgements.--}LGDS acknowledges support from Brazilian agencies
FAPESP (2013/50220-7), CNPq (307107/2013-2) and PRP-USP NAP-QNano. We acknowledge useful discussions with M.~Le~Dall, E. Miranda, K. Ingersent, H.~E. T\"{u}reci and I. \v{Z}uti\'{c}, and financial support from the Canadian program NSERC-Discovery
and a generous FAPESP-UVic exchange award.
\end{acknowledgements}

\appendix

\section{Details of the NRG calculations}
\label{sec:NRGdetails}

As we argue in the main text, our choice of the Complete Fock Space (CFS) procedure \cite{Peters:Phys.Rev.B:245114:2006} (or, equivalently, the full density matrix NRG method (FDM-NRG)\cite{Weichselbaum:Phys.Rev.Lett.:99:076402:2007} at $T\!=\!0$) in the NRG calculations presents some advantages for the calculations of the correlation functions listed in Eq.\ (\ref{Eq:noise_Lehmann_delta}). To illustrate this point, we compared results obtained using CFS and the earlier ``Density Matrix-NRG" (DM-NRG) method.\cite{Hofstetter:1508:2000} 

\begin{figure}[t]
\includegraphics[width=0.5\textwidth]{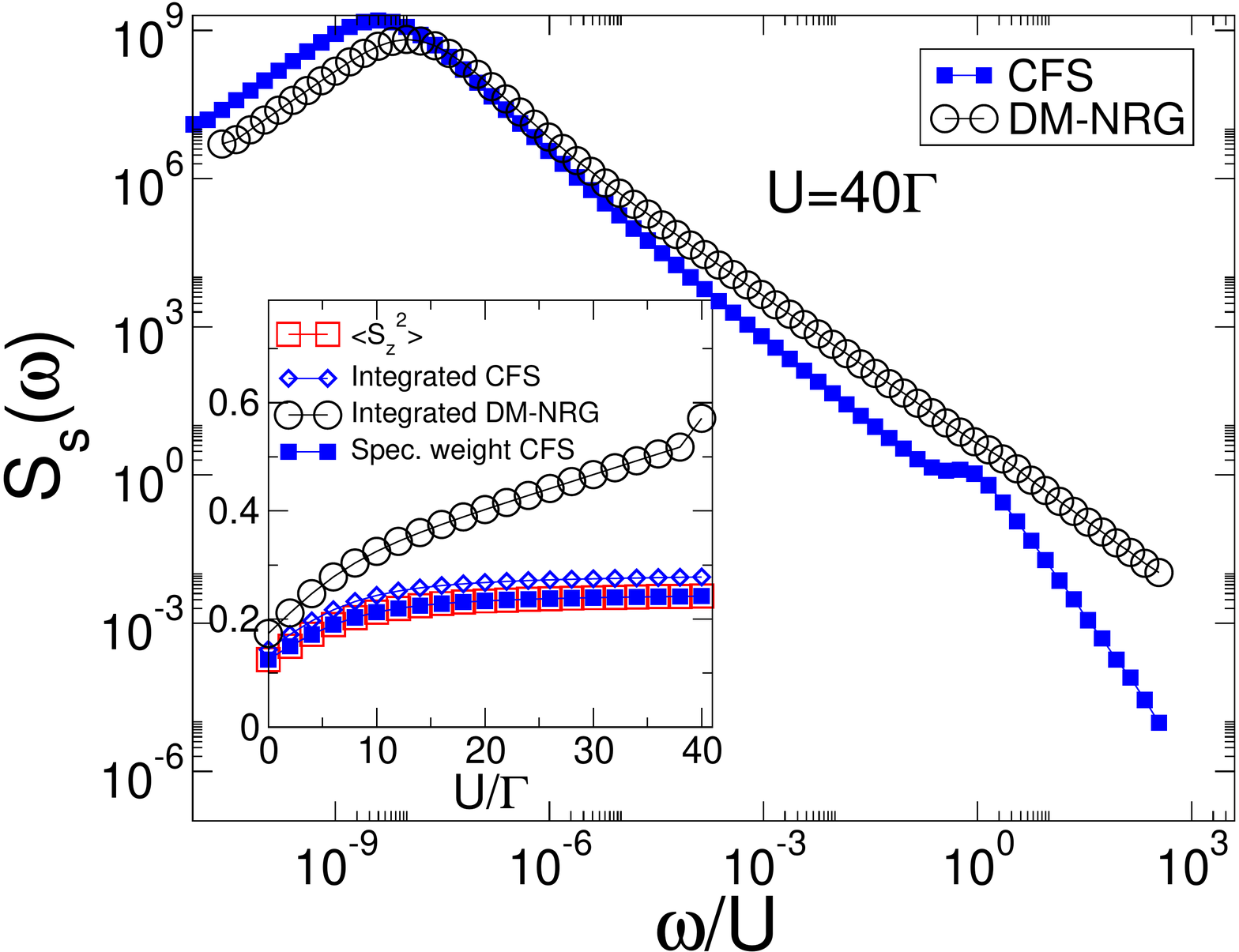}
\caption{(Color online) NRG data for the spin noise $S_s(\omega)$ at $T\!=\!0$ for $U=40 \Gamma$ calculated with DM-NRG and CFS procedures. Inset: A check of the spin sum rule given by Eq.\ \ref{Eq:SumRule} for the different approaches shows that the CFS fulfils the spin sum rule apart from numerical integration errors.}
\label{fig:DMNRG_CFS_SumRule}
\end{figure}

The main panel in Fig.\ \ref{fig:DMNRG_CFS_SumRule} presents NRG data for the spin noise $S_s(\omega)$ using DM-NRG (open circles) and CFS (filled squares) for $U=40 \Gamma$ and other parameters set as in Fig.\ \ref{fig3}. In both cases, the NRG calculations were performed using a discretization parameter $\Lambda=2.5$ retaining up to 1000 states at each NRG step, which ensures convergence for the single-trap Anderson model. The spectral data was broadened using the usual logarithmic Gaussian functions (Eq. (74) in Ref.\ \onlinecite{Bulla:395:2008}) with a broadening parameter $b=\log(\sqrt{\Lambda})\approx 0.46$ (We have used z-averaging for some of the data presented, particularly the data presented in Fig.\ \ref{fig1}).

Clearly, DM-NRG subestimates the peak at $\omega\!=\!T_K$ in comparison with CFS. More importantly, it misses the transition from the $S_s(T_K \! \ll \!\omega\! \ll \!U) \propto
1/[\omega\log^{2}{(\omega/T_K)}]$ to $S_s(\omega \gg U) \propto \omega^{-2}$ behaviors, which is one of the important features distinguishing the spin noise from the charge noise.

We have also compared both methods by checking the fulfillment of the spin sum rule in Eq.~(\ref{sumrule1}). In the absence of spin polarization (due to, e.g. magnetic fields or ferromagnetic couplings), $\langle\hat{S}_z \rangle\!=\!0$ and the spin sum rule is given by: 
\begin{equation}
\label{Eq:SumRule}
\int_{-\infty}^{\infty}S_s(\omega) \; d \omega \! = \!  \langle\hat{S}^2_z \rangle.
\end{equation}

To this end, we performed a static NRG calculation of $\langle\hat{S}^2_z \rangle (T \rightarrow 0)$ (open squares in the inset of Fig.\ \ref{fig:DMNRG_CFS_SumRule}) and compared with a numerical integral of $S_s(\omega)$. The agreement of the integrated CFS data (diamonds) is much better than the DM-NRG (circles), although the fulfillment of the spin sum rule is not perfect due to numerical errors coming from the integration procedure.

Alternatively, the integral in Eq.\ (\ref{Eq:SumRule}) can be done directly in Eq.\ (\ref{Eq:noise_Lehmann_delta}) and it becomes a sum over the spectral weights $|\langle n|\hat{S}_z|m \rangle|^2$ provided that the set of many-body states $\{|m \rangle\}$ forms a complete set. In practice, this procedure can be done in the CFS scheme, as it retains matrix elements between ``discarded" and ``kept" NRG states, making the set of many-body states complete by construction.\cite{Peters:Phys.Rev.B:245114:2006,Weichselbaum:Phys.Rev.Lett.:99:076402:2007} In this case, free of numerical integration errors, the CFS data (filled squares) fulfills the spin sum rule down to machine precision, as shown in the inset of of Fig.\ \ref{fig:DMNRG_CFS_SumRule}. 

\section{Validation of Hartree-Fock approximation for charge noise when $T\geq 0$}
\label{sec:apphf}

The HF approximation Eq.~(\ref{mftc}) was shown to approximate charge noise at $T=0$. Here we check its validity at $T\geq 0$, by direct evaluation of the Shiba relations and sum rules described in Section~\ref{sec:Model}. 

The static ($\omega=0$) charge susceptibility in the HF approximation is given by
\begin{eqnarray}
\chi_{c}^{HF}(\omega=0,T)&=&\frac{1}{2\pi}\frac{\partial}{\partial\epsilon_F}\sum_\sigma\langle n_{\sigma}\rangle \nonumber\\
&=& \frac{1}{2\pi}\int d\epsilon \sum_\sigma A_{\sigma\sigma}(\epsilon) \frac{\partial f(\epsilon)}{\partial\epsilon_F}\nonumber\\
&=& \frac{1}{8\pi T} \int d\epsilon \frac{\sum_\sigma A_{\sigma\sigma}(\epsilon)}{\cosh^{2}\left(\frac{\epsilon-\epsilon_F}{2T}\right)}.\label{chichf}
\end{eqnarray}
At $T=0$ we get 
\begin{equation}
\chi_{c}^{HF}(\omega=0,T=0)=\frac{1}{2\pi}\sum_\sigma A_{\sigma\sigma}(\epsilon_F).
\end{equation}

We start by checking the Shiba relation for charge noise, Eq.~(\ref{shibacharge}). In the HF approximation we get
\begin{equation}
{\rm Lim}_{\omega\rightarrow 0^+} \int_{\epsilon_F}^{\epsilon_F+\omega}\frac{d\epsilon}{\omega} \sum_\sigma A_{\sigma\sigma}(\epsilon)A_{\sigma\sigma}(\epsilon-\omega)=\sum_\sigma A_{\sigma\sigma}^{2}(\epsilon_F),
\label{shibaHFL}
\end{equation}
which according to the Shiba relation should be equal to
\begin{equation}
2\pi^2 [\chi_{c}^{HF}(0,0)]^2 = \frac{1}{2}\left[\sum_\sigma A_{\sigma\sigma}(\epsilon_F)\right]^{2}.
\label{shibaHFR}
\end{equation}
The relation is satisfied exactly at $U=0$; however, as $U$ increases Eq.~(\ref{shibaHFL}) becomes up to two times larger than Eq.~(\ref{shibaHFR}). This discrepancy can indeed be observed in the comparison with NRG, see the difference in slopes at $\omega=0$ in Fig.~\ref{fig1}. Nevertheless, the discrepancy is not too large.  

The charge sum rule [Eq.~(\ref{sumrule1}) for $i=c$] in the HF approximation reads
\begin{eqnarray}
\int_{-\infty}^{\infty} d\omega S_{c}^{HF}(\omega,T) &=& \sum_\sigma \int d\epsilon A_{\sigma\sigma}(\epsilon)[1-f(\epsilon)]\nonumber\\
&&\times\int d\omega A_{\sigma\sigma}(\epsilon-\omega)f(\epsilon-\omega)\nonumber\\
&=& \sum_\sigma [1-\langle n_{\sigma}\rangle]\langle n_{\sigma}\rangle\nonumber\\
&=& \langle {\cal
\hat{O}}^2_c \rangle_{HF} - \langle {\cal \hat{O}}_c \rangle_{HF}^{2},
\label{sumrule1HF}
\end{eqnarray}
where
$\langle {\cal \hat{O}}_c \rangle_{HF}=\sum_\sigma \int d\epsilon
A_{\sigma\sigma}(\epsilon)f(\epsilon)$
and $\langle {\cal \hat{O}}^2_c \rangle_{HF}$ is obtained by making
the approximation
$\langle n_{\uparrow}n_{\downarrow}\rangle \approx\langle
n_{\uparrow}\rangle\langle n_{\downarrow}\rangle$.
The last line of Eq.~(\ref{sumrule1HF}) is
expected to be a good approximation to the exact result, even in the
Kondo regime, when charge fluctuations are strongly suppressed. 

Finally, we verify the charge susceptibility sum rule [Eq.~(\ref{sumrule2}) for $i=c$] with explicit numerical calculations of the quantity
\begin{equation}
{\rm Sum}_{\chi_c}=\frac{1}{\chi_{c}^{HF}(0,T)} \int_{-\infty}^{\infty}d\omega \frac{1-\textrm{e}^{-\omega/T}}{2\pi \omega}S_{c}^{HF}(\omega,T).\label{sumchic}
\end{equation}
As shown in Table~\ref{tablechargecheck} these values are very close to $1$ for all tested parameters. 

In conclusion, the HF approximation for charge noise is consistent with the exact relations of Section~\ref{sec:Model} for all parameters checked, indicating that it provides a good analytical approximation for charge noise even for $T>0$.

\begin{table} 
\begin{center} 
\begin{tabular}{c |c |c | c } 
$T/\Gamma$ & $\epsilon_d/\Gamma$ & $U/\Gamma$ & ${\rm Sum}_{\chi_c}$\\
\hline 
$0.1,\;1,\;10$  &  $0$  & $0$ & $0.9997,\;0.9999,\;1.000$\\ 
$0.1,\;1,\;10$  &  $-2.5$  & $5$ & $0.9994,\;0.9999,\;1.000$ \\ 
$0.1,\;1,\;10$  &  $-10$  & $20$ & $0.9915,\;0.9978,\;1.000$ \\ 
$0.1,\;1,\;10$  &  $0$  & $5$ & $0.9996,\;0.9999,\;1.000$ \\ 
$0.1,\;1,\;10$  &  $-10$  & $10$ & $0.9996,\;0.9999,\;1.000$ \\ 
\hline 
\end{tabular} 
\caption{Charge susceptibility sum rule in the HF approximation [Eq.~(\ref{sumchic})]. The sum rule is seen to be 
satisfied (${\rm Sum}_{\chi_c}=1$) with high accuracy for several different parameters.} 
\label{tablechargecheck} 
\end{center} 
\end{table}


\end{document}